\documentclass[12pt,preprint]{aastex}

\def\kms{\relax \ifmmode {\,\rm km~s}^{-1}\else \,km~s$^{-1}$\fi}
\def\cm3{cm$^{-3}$}
\def\Jb{\relax \ifmmode {\,\rm Jy\,beam}^{-1}\else \,Jy\,beam$^{-1}$\fi}
\def\mJb{\relax \ifmmode {\,\rm mJy\,beam}^{-1}\else \,mJy\,beam$^{-1}$\fi}
\def\deg{\hbox{$^\circ$}}
\def\arcmin{\hbox{$^\prime$}}
\def\arcsec{\hbox{$^{\prime\prime}$}}
\def\secd#1.#2{ #1\farcs#2 }               % seconds over decimal point
\def\e{$\pm$}
\def\x{$\times$}
\def\j21{$J$=2$\rightarrow$1}
\def\c3h2{C$_3$H$_2$}
\def\nh3{NH$_3$}
\def\n2h+{N$_2$H$^+$}

\slugcomment{To appear in the Astronomical Journal}

\begin{document}

\title{The Physical and Chemical Status of Pre-protostellar Core B68}
\author{Shih-Ping Lai, T.\ Velusamy, W.\ D.\ Langer, and T.\ B.\ H.\ Kuiper}
\affil{MS 169-506, Jet Propulsion Laboratory, California Institute of 
Technology, \\ 
Pasadena, CA 91109; slai@pangu.jpl.nasa.gov}

\begin{abstract} 

We have investigated the physical and chemical status of 
the pre-protostellar core B68.  A previous extinction study suggested that
the density profile of B68 is remarkably consistent with
a Bonnor-Ebert sphere with 2.1 $M_{\odot}$ at 16 K.
We mapped B68 in \c3h2, CCS, and \nh3 with 
the Deep Space Network (DSN) 70m telescope at Goldstone.
Our results show that the \nh3 peak coincides with the dust continuum peak,
whereas CCS and \c3h2 are offset from the \nh3 and dust  peaks.
The B68 chemical structure is consistent with that seen in other such 
pre-protostellar cores (L1498, L1544) and is explained 
by time dependent 
chemical models that include depletion.
We measured the kinetic temperature of B68 with \nh3 (1,1) and
(2,2) spectra obtained with a DSN 34m telescope.
We find that the kinetic temperature of B68 is 
only 11 K which is significantly lower than that previously assumed.   
We also derive the non-thermal linewidth in B68,
and show that B68 is thermally dominated with little contribution from 
turbulence support ($<$ 10\%).   
We consider a modified Bonnor-Ebert sphere to include effects of turbulence and 
magnetic fields and use it to constrain the uncertainties in its distance 
determination. 
 We conclude that 
the distance to B68 is $\sim$ 95pc
with a corresponding mass of $\sim$ 1.0 $M_{\odot}$.
If some  magnetic field
is present it can be   further away  (beyond $\sim$ 100pc) and still 
satisfy the density structure of a Bonnor-Ebert sphere.   The sulfur (CS and 
CCS) and carbon chain (\c3h2)  molecules are heavily depleted in B68 and do not 
trace the dense interior region. We see some  evidence for 
depletion of \nh3 at the core center roughly on a  scale similar to that of 
\n2h+. 
Our observations do not preclude any instability   such as the onset  of  
collapse, or slow contraction,  
occurring in the center of the core, 
which cannot be resolved with our beam size (45$^{\prime\prime}$).
\end{abstract}


\section{Introduction}

What are the initial conditions that give rise to star formation?
Understanding the physical and chemical structure of the cores
at the pre-collapse stage (``pre-protostellar cores")
is critical for examining models for the core evolution 
and star formation.  Several theoretical models have been proposed 
for pre-protostellar cores; recently one of these, the Bonnor-Ebert sphere, 
has been the subject of increased interest.
A pre-protostellar core may be described as a hydrostatic 
isothermal pressure-confined, non-rotating, non-magnetized,
self-gravitating sphere -- a Bonnor-Ebert sphere (Bonnor 1956; Ebert 1955).  
If the center to edge density contrast of a Bonnor-Ebert sphere 
exceeds a critical value, this core is unstable and subject to 
collapse with small perturbation.
The extreme case of such collapsing cores is a singular isothermal sphere
with density $n \propto r^{-2}$, which will collapse in a free-fall time
scale (Shu 1977). Recent extinction study by Alves, Lada, \& Lada  
(2001; hereafter ALL2001) suggested that
the density profile of B68 is remarkably consistent with
a Bonnor-Ebert sphere with 2.1 $M_{\odot}$ at 16 K.
As turbulence exists prevalently in giant molecular clouds,
a Bonnor-Ebert sphere must be formed after turbulence has decayed;
however, it is not well understood how this transition occurs
and how long it takes for the decay of turbulence.
It is, therefore, important to determine 
observationally the degree of turbulence present in B68.   Furthermore,
the presence of magnetic fields in some pre-protostellar cores have been
detected through dust polarization (Ward-Thompson et al.\ 2000).
If magnetic fields couple to the gas dynamics through the ionized gas
component (ambipolar diffusion), 
magnetic fields could slow down the dynamical collapse 
of the cores (Ciolek \& Mouschovias 1993).
These models predict a different physical structure for the density, 
temperature, magnetic fields, and turbulence,
which will determine the  dynamics and evolutionary history of the core.
Both turbulence and magnetic fields can provide additional support against 
gravity. Thus, for a  given/observed center-to-edge density contrast, the 
presence of any turbulence and magnetic field  can modify the
size and hence the mass, for an equivalent Bonnor-Ebert sphere.  

The density structure of pre-protostellar cores is obtained through
three major approaches:
(sub-)millimeter and infrared dust emission (e.g. Shirley et al.\ 2000;
Willacy \& Langer 2000), dust extinction from optical and infrared 
wavelengths (e.g. ALL2001), and molecular gas emission 
with tracers that do not readily deplete at high density such as \nh3 and 
\n2h+ (e.g. Kuiper et al. 1996; Tafalla et al. 1998).
However there are limitations to each of these probes of the core status. 
The dust is optically thin at submillimeter and far infrared wavelengths and
thus can be well mapped.  
However, the interpretation of the dust emission is sensitive to  both 
temperature and  density. 
The dust extinction obtained with optical and infrared has
very high spatial resolution; however, it requires  background stars 
along each line of sight across the core
and the derivation of density relies on the assumed gas-to-dust ratio.
Nevertheless when a grid of a large number of stars are observed, as 
demonstrated by ALL2001 for B68,  it can provide a robust 
estimate of the core size and the relative density profile across the core.
The molecular spectral line observations are important probes of core 
dynamics, which cannot be obtained with dust observations.
However they too have limitations as density and dynamical probes because of 
depletion.
Chemical models suggest that most molecules may deplete   
in the center of the cores as molecules accrete onto dust grains
(Rawlings et al.\ 1992; Bergin \& Langer 1997), 
making H$_2$ column density and 
mass determinations subject to the assumed depletion factor.
NH$_3$ and N$_2$H$^+$ are among the very few molecules
whose abundances remain high
even when others, e.g., CO, CCS and CS, are heavily depleted.

Submillimeter and millimeter continuum maps of pre-protostellar cores have 
shown that the typical intensity profile of pre-protostellar cores is
flat at the inner region and steepens toward the edge
(Ward-Thompson et al.\ 1994; Ward-Thompson, Motte, \& Andre 1999, 
Willacy \& Langer 2000).  
If the temperature gradient in the cores is negligible, 
the intensity profile can be treated as a column density profile.
Such profiles are consistent with the predicted signature of 
Bonnor-Ebert spheres (Ward-Thompson 2002), but they are also
qualitatively consistent with the ambipolar diffusion model 
(Ciolek \& Basu 2000).
 
B68 is an isolated quiescent Bok Globule near the Ophiucus complex.
Its nearly spherical geometry makes it a good candidate for comparison 
between observations and theoretical models 
(which are primarily spherically symmetric).
The importance of B68 is that it may be a core on the verge of 
collapse. ALL2001 derive the radial extinction profile of B68 from 
the optical and 
near infrared photometry of background stars, and 
the profile is remarkably 
consistent with a 16 K isothermal Bonnor-Ebert sphere
with a mass of 2.1 M$_\odot$ (assuming a distance of 125pc) 
which is slightly over the critical mass.
Their result  on the center-to-edge density contrast implies that B68 can 
adequately be supported by the thermal
pressure without the support from turbulence and/or magnetic fields
before the dynamical collapse takes place.
As the best candidate of a Bonnor-Ebert sphere to date,
B68 has attracted great attention recently.
Langer \& Willacy (2001) used the ISO 200$\mu$m  and 160$\mu$m data and obtained 
a density profile that can be fitted by a series of power laws 
with exponents -1.2, -2, and -4, going from center to edge,
which is similar to the density profile of other pre-protostellar cores.   
Recently the submillimeter dust continuum has been mapped at 450 and 850 
$\mu$m with SCUBA (Visser, Richer, \& Chandler 2002).  
However, to date, the extinction 
data of ALL2001 has the highest spatial resolution (3\arcsec) and  provides a 
robust 
density profile and estimate for its angular size. 

Knowledge of the distance to the core (to estimate the size and mass ) and 
temperature (to estimate the thermal pressure)
are critical to characterize the true status of B68 as a prestellar core. 
Hotzel et al.\ (2002) obtain T$\sim$8K from the excitation analysis 
of CO lines and suggest that the previous temperature measurement from
Bourke et al.\ (1995) contains calibration or calculation errors.
In a later paper, Hotzel, Harju, \& Juvela (2002) also derived
a kinetic temperature of 10$\pm$1.2 K from the less depleted \nh3.
Here, we present  spectral line maps of B68 in \c3h2, CCS, and \nh3 
to characterize its   chemical status
and to derive the temperature and 
turbulence in the core. Our results when combined with the   density 
structure derived from extinction data (ALL2001) provide stringent constraints
on the distance (and thus, linear size and mass) to the B68 core.

\section{Observations }

We made single-dish maps of B68 in \c3h2, CCS, and \nh3 lines between  
January and July 2002  using NASA's Deep Space Network (DSN) 70 m antenna 
at Goldstone.
We mapped B68 over $\sim$4\arcmin\x4\arcmin\ region with Nyquist sampling 
every 25\arcsec\ around the nominal center position from Benson \& Myers (1989)
at R.A.(1950)=17$^h$19$^m$36$^s$, Dec.(1950)=$-$23\deg47\arcmin13\arcsec.
The maps were made with a K-band HEMT receiver tunable between 18--26 GHz.
The spectra of these lines were taken with the two-million-channel
Wide Band Spectral Analyzer (WBSA; Quirk et al.\ 1988).  
We co-added 256 adjacent channels   in real time to produce a 8192 point 
spectrum with 5 kHz spectral resolution and high signal-to-noise ratio.
We also used a DSN 34m antenna and the WBSA to obtain the spectra of \nh3 (1,1) 
and (2,2) lines simultaneously 
in order to derive the kinetic temperature of B68. 
The spectrum were taken at (0\farcm25, 0\farcm25) offset from the nominal 
center of B68 so that the 34 m beam includes most of the \nh3 emission.
A CCS $J_N=2_1\rightarrow 1_0$ spectrum was also obtained with the DSN 34m 
at (0\farcm25, 0\farcm25) to determine the non-thermal linewidth.
In Table 1 we list the observed transitions and their spatial and spectral 
resolutions.

\section{Results}

\subsection{Chemical Differentiation}

Figure 1 shows our DSN 70m maps of \c3h2, \nh3, and CCS.
The \nh3 map shows a centrally peaked distribution with
the maximum close to the nominal center.  By contrast,
\c3h2 and CCS both have their strongest peaks
to the northeast of the \nh3 peak and their intensity decreases
towards the southwest of the \nh3 peak. The \c3h2 map shows a low 
level emission along the southwest edge.
CCS has another strong peak at the southeast extension of B68.   
The JCMT dust continuum maps of B68 at 850$\mu$m 
(Visser, Richer, \& Chandler 2002) and CS emission (Lada et al.\ 2003)
also show a distinct peak at the southeast extension, 
indicating this extension could be a separate core.

In Figure 2 we show an overlay of the \nh3 map with the \n2h+ map 
obtained at the IRAM 30m telescope (Bergin et al.\ 2002) and 
the 200~$\mu$m dust emission obtained with ISO (Langer \& Willacy 2001).
We found that the \nh3 peak coincides with 
the dust continuum peak and is located between the two \n2h+ peaks in the 
IRAM  map.
This result is consistent with the expectation of the chemical models
that \nh3 and \n2h+ are relatively more abundant in high density 
regions than other molecules.  The double peak of \n2h+ has been
interpreted to indicate that even \n2h+ depletes in the region with 
$A_v>25$mag.  
A similar depletion of \nh3 could take place in the center of B68,
but is not spatially resolved with our observations.

\subsection{Kinetic Temperature}

At low temperatures, where collisional excitation dominates, the rotation 
temperature of \nh3 can be estimated from the intensity ratio of the (1,1) 
and (2,2) lines $T_{22}/T_{11}$ and the optical depth of the (1,1) 
main line $\tau_{11}$ (Ho \& Townes 1983):
\begin{equation}
T_{r}=-41.5/\ln\left(\frac{-0.282}{\tau_{11}} 
\ln\left(1-\frac{T_{22}}{T_{11}} 
\times (1-e^{-\tau_{11}})\right)\right)
\end{equation}
\noindent We fitted our 34m \nh3 (1,1) and (2,2) hyperfine lines using 
the Continuum and Line Analysis Simple Software (CLASS), and
obtained $T_{22}/T_{11}=0.20\pm0.04$K and $\tau_{11}=2.5\pm0.2$.
From Eq(1), we derived $T_{r}=10.9\pm0.8$K (Lai et al. 2002).
Figure 4 show how the uncertainty in the estimate of $T_{r}$ 
varies with $T_{22}/T_{11}$ and $\tau_{11}$.
The rotation temperature slightly underestimates the kinetic temperature
$T_k$ in dark clouds (Walmsley \& Ungerechts 1983),
but it approaches the kinetic temperature when $T_k\rightarrow 10K$.
We adopt $T_k=11K$ in further calculations.
Independently, Hotzel, Harju, \& Juvela (2002)
also measured a kinetic temperature of 10$\pm$1.2K in B68 from \nh3 
observations using the Effelsberg 100m telescope.

\subsection{Non-thermal linewidth}

The non-thermal linewidth (FWHM), $\Delta v_{nth}$, can be estimated from
$\Delta v_{obs}^2 = \Delta v_{th}^2 + \Delta v_{nth}^2$.
The observed linewidth $\Delta v_{obs}$ can be obtained from
the spectral line data. The thermal linewidth $\Delta v_{th}
=\sqrt{(8\ln 2) k T_k / \mu m_H}$  ($\mu$ is the atomic weight
and $m_H$ is the mass of a hydrogen atom) can be calculated with
the kinetic temperature determined in \S3.2.
We derived $\Delta v_{nth}$ from \nh3 and CCS separately.
The multiple line fitting of \nh3 gives a reliable determination of 
the observed linewidth.  The CCS molecule has a large atomic weight,
and hence a narrower thermal linewidth.
We derived for both \nh3 and CCS  $\Delta v_{nth}\sim$0.14\kms\ (Table 2).
These values are consistent with that   
observed from  C$^{18}$O line widths ($\Delta v_{nth}\sim 0.18$\kms) 
by Lada et al.\ (2003).
The $\Delta v_{nth}$, interpreted as that due to turbulence, 
is small -- of the order of 0.3$\times \Delta v_{th}$(H$_2$).

\subsection{Column Density}

We calculate the column density of \nh3, CCS, and \c3h2
under the assumption of LTE at two positions, one at
the \nh3 peak and one at the CCS peak, in order to compare
our results to the chemical evolution models.
The derived column densities are listed in Table 3.

The \nh3(1,1) hyperfine lines allow us to obtain the antenna temperature
$T_A$ and the optical depth $\tau$ directly; therefore, we can
derive the excitation temperature $T_{ex}$ of \nh3
from the radiative transfer equation
\begin{equation}
T_B = T_A/\eta=(T_{ex}-T_{bg})(1-e^{-\tau}),
\end{equation}
\noindent where $T_B$ is the brightness temperature,
$T_{bg}=2.7K$ is the cosmic background temperature, and
$\eta\sim$0.7 is the main beam efficiency of the DSN 70 m antenna.
The column density of \nh3 can be estimated with the knowledge of
the excitation temperature $T_{ex}$,
the line width $\Delta v$ and the main group opacity of the (1,1) line
$\tau_{11}$ by the usual assumption that only metastable levels are
populated (Harju, Walmsley, \& Wouterloot 1993), therefore
\begin{equation}
N(NH_3) = 1.6 \times 10^{13} \Delta v~\tau_{11}~\frac{\exp(h\nu
/kT_{ex})+1}{\exp(h\nu /kT_{ex})-1}~(\frac{1}{3} e^{23.4/T_r}+1+\frac{5}{3}
e^{-41.5/T_r}+\frac{14}{3} e^{-101.5/T_r}+...)~~~{\rm cm^{-2}}
\end{equation}
\noindent where $T_r$ is the rotation temperature derived in \S3.2.
We assume $T_r$ is constant across the core.

The column density of CCS can be derived from
\begin{equation}
N(CCS)=5.1\times 10^{11}~\tau~\Delta
v~U~\frac{\exp(E_u/kT_{ex})}{\exp(h\nu/kT_{ex}-1)}~~~{\rm cm^{-2}}
\end{equation}
\noindent where $E_u$ is the energy level of the upper state and $U$ is
the partition function (Suzuki et al.\ 1992).
The excitation temperature of CCS is unknown, because we only have one
transition.  Wolkovitch et al. (1997) has done excitation analysis
with three CCS transitions for L1498 and TMC-1D, and the kinetic
temperature for these two cores is 7--10K.
Suzuki et al.\ (1992) derive an   excitation temperature of 5 K
for a number of quiescent dark cores. Therefore, we
assume $T_{ex} \sim 5-10$~K and the partition function $U$ is 24-62
for the lowest 76 rotational transitions.

For \c3h2 we use the approach of Bell et al. (1988),
\begin{equation}
N(C_3H_2) = 1.98\times 10^{-20}~\frac{\tau~\nu^2~\Delta\nu~Q~\exp(E_J/kT_{ex})}{
A_{J'J}~g_{J'J}~[1-\exp(-h\nu/kT_{ex})]}~~~{\rm cm^{-2}}
\end{equation}
\noindent where $\nu$ is the frequency of the line in Hz, $\Delta\nu$ is
the FWHM in Hz, $E_J=2.352K$ is  the energy of lower level,
and the Einstein constant $A_{J'J}=4.02\times10^{-7}~s^{-1}$.
The partition function Q is calculated from
$Q=(5.34/\sigma)\times(T_{ex}^3/ABC)^{1/2}$, where
A=35.092596 GHz, B=32.212931 GHz, C=16.749315 GHz, and $\sigma=2$
for \c3h2 (Thaddeus, Vrtilek, \& Gottlieb 1985).
We also assume the excitation temperature is in the range of 5--10K.

\section{Discussion}
\subsection{Modified Bonnor-Ebert Sphere}

A Bonnor-Ebert sphere is a hydrostatic self-gravitating isothermal core 
solely supported by its thermal pressure.  The density profile of
a Bonnor-Ebert sphere is characterized by only one parameter,
the dimensionless radius parameter $\xi_{max}$,
\begin{equation} 
\xi_{max}=(R/a)\sqrt{4\pi G\rho_c},
\end{equation}
where R is the core radius, $\rho_c=m n_c$ is the central density, 
$a=\sqrt{kT/m}$
is the isothermal sound speed, and $m$ is the mean molecular weight.
ALL2001 demonstrate that the density profile of B68 inferred from 
extinction is remarkably 
consistent with that of a Bonnor-Ebert sphere with $\xi_{max}$=6.9.
However   the results of our observations   show that (1) B68 is colder than
what ALL2001 assumed ($T_k$=11K rather than 16K),
and (2) thermal support dominates over turbulence, but that turbulent 
energy is present.   The lower temperature will only affect 
the scaling of the density profile, but not 
the interpretation of the nature of B68 as a Bonner-Ebert sphere. 
We discuss how our observations constrain the parameters in $\xi_{max}$, and 
hence the distance and the mass
of B68. We also consider an empirically modified Bonnor-Ebert sphere that includes turbulent 
and magnetic support. 

The density profile of the Bonnor-Ebert sphere can be used to constrain
the distance and the mass of B68 with a temperature measurement.
The distance to B68 itself is not well determined.
B68 is spatially close to the Ophiuchus complex, and
the distance to the Ophiuchus complex is between 80 and 170 pc with
a central value of 125 pc (de Geus et al.\ 1989). 
ALL2001 have adopted 125pc in their modeling. 
Hotzel et al.\ (2002) have presented detailed discussion 
on the physical quantities of a Bonnor-Ebert sphere and 
have derived a mass of 0.7 $M_\odot$ and a distance of 80pc
using $T_k=8$ K from CO observations.  
Here we shorten the derivation of Hotzel et al.\ (2002) and 
directly calculate the distance 
and the mass of B68, using dimensional analysis of the three observed
quantities from the extinction data 
which are invariant to temperature $T$ and distance $D$:
the dimensionless radius parameter $\xi_{max}$, the angular diameter
$\theta$, and the the central column density $N_c$. We can express 
\begin{equation} 
\frac{\xi_{max}}{\theta}=m~D~\sqrt{\frac{4\pi G n_c}{k T}}.
\end{equation} 
\noindent Because $\xi_{max}$ and $\theta$ are fixed by observations, 
from Eq(7) 
the number density at the center $n_c \propto D^{-2}~T$.   Furthermore, 
$n_c$ is also related to the column density $N_c=\int n~ds= K n_c D \theta$, 
where K is a constant which depends only on
the shape of the density profile characterized by $\xi_{max}$.
Because the central column density $N_c$ is fixed by observation,
\begin{equation} 
n_c = \frac{N_c}{K~D~\theta} \propto D^{-1}.
\end{equation} 
Since $n_c \propto D^{-2}~T$ and also $\propto D^{-1}$, we find
$D^{-1}~T$ is a constant.
Therefore, the temperature measurement can be used to constrain
the distance.  ALL2001 assumed D=125pc and T=16 K, 
to derive a model consistent with $\xi_{max}=6.9$ and 
$M=2.1~M_\odot$. With our measured temperature of 11 K,
B68 would have to be at a closer distance $\sim$ 85 pc with a mass of 
$\sim$ 1.0 $M_\odot$ for the same $\xi_{max}$. 
This would place B68 on the near side of the Ophiuchus complex.

The isothermal Bonnor-Ebert sphere is supported purely by thermal 
pressure and does not include effects of turbulence or magnetic fields. 
Turbulence and magnetic fields have
been observed in some pre-protostellar cores and therefore they may provide 
significant support to the core stability. We present an empirical modification 
to the Bonnor-Ebert sphere that includes the  turbulence and magnetic field
support while preserving a Bonner-Ebert sphere-like density profile consistent 
with the observed $\xi_{max}$. Though empirical, such a representation is 
useful for a qualitative study of the core properties.

The observed turbulence  in B68 (reported in this paper) provides 
additional support to the core.  We can incorporate the turbulent support by   
replacing the temperature in Eq(7)
with an effective temperature $T_{eff}$ that includes both
thermal  and the turbulent pressures:   
$T_{eff}=T_k+T_{turb}$, where $T_{turb}$ is obtained from the turbulence 
velocity width, 
$\Delta v^2_{nth}=(8\ln2)k T_{turb}/m$.   We can now modify  Eq(7) as
\begin{equation} 
\frac{\xi_{max}}{\theta}=m~D~\sqrt{\frac{4\pi G n_c}{kT_k + \frac{m \Delta 
v_{nth}^2}{8\ln2}}}.
\end{equation} 
\noindent We use Eqs(8) \& (9) and the observed $T_k$ and  $\Delta v^2_{nth}$  
to constrain the distance of B68 to be 85--100pc, as shown in Figure 5a.

Eq(9) can be modified further to include a magnetic 
pressure term. However, in order to preserve the Bonner-Ebert-like density 
profile, this modification is possible only in cases where the magnetic 
pressure also varies 
radially as the thermal pressure. That is, magnetic pressure
 $P(r)\sim n(r)$.
Coincidentally, theoretical models of an axially symmetric isothermal core 
with magnetic flux freezing give $B \sim n^{1/2}$ 
(Fiedler \& Mouschovias 1993), 
therefore magnetic pressure $B^2/8\pi \sim n$.
Zeeman observations in molecular clouds also support such a  scaling law
(Crutcher 1999).
Therefore, we can rewrite Eq(9) to include magnetic field 
support with $T_{eff}=T_k+T_{turb}+T_{mag}$, where $T_{mag}$ is the equivalent 
temperature corresponding to the magnetic pressure, $B^2$.
For simplicity, we express $T_{mag}$ in terms of magnetic pressure using the 
average magnetic field $\bar{B}$
and the average density $\bar{n}$. 
We obtain the following relation between $\xi_{max}$ and the kinetic 
temperature ($T$), distance $D$, central density n$_c$, 
turbulent velocity width $v_{nth}$, and magnetic field strength $B$.
\begin{equation} 
\frac{\xi_{max}}{\theta}=m~D~\sqrt{\frac{4\pi G n_c}{kT + \frac{m \Delta
v_{nth}^2}{8\ln2}+\frac{\bar{B}^2}{12\pi \bar{n}}}}.
\end{equation} 

Though Eq(10) is valid only under certain conditions (when $B$ 
scales as $n^{1/2}$) it is useful to show the dependence of the observed 
parameters on the physical conditions in the core. 
We use Eq(10) to constrain the linear size (or its distance)
for B68 as a function of kinetic temperature, turbulence,
and magnetic field strengths as shown in Figure 5.
For any given choice of turbulence with no magnetic fields,
the distance and temperature along the lines shown in Figure 5a represent 
stable cores that can be described with Bonner-Ebert density profiles,
while those above and below represent supercritical 
and subcritical cores respectively. 
The shaded area in Figure 5a marks the 
best estimate for distance consistent with our measurements of 
temperature and turbulence when magnetic support is ignored. 
Figure 5b shows the distance, for which B68 will be static, 
as a function of increasing magnetic field derived using Eqs(10) \& (8).
(The temperature and turbulence widths are fixed at the observed value 
of 11K and $\Delta v_{nth} = 0.3 \Delta v_{th}$.)
Because $n_c$ and $\bar{n}$ are both proportional to $D^{-1}$,
as long as $\xi_{max}$ and $N_c$ are fixed, 
the distance can be derived for any given $\bar{B}$.
Figure 5b indicates that if B68 is at distances beyond $\sim$100pc, 
then it can  
remain a stable core only if magnetic fields are present. 
However, under the flux freezing condition, the magnetic field
should be responsible for the excitation of the non-thermal linewidths.
The observed 0.14\kms\ non-thermal linewidths only gives $B\sim 5-10\mu$G
for density of 1--3$\times 10^4$ \cm3.

We conclude that the Bonner-Ebert-like density profile does not necessarily
rule out the existence of the turbulence and magnetic fields,
although their contributions are small in the case of B68.
The fact that B68 is dominated largely by thermal support suggests that
the turbulence and the magnetic flux have been dissipated
and the core could be on the verge of the collapse.
The observed $\xi_{max}=6.9$ is slightly larger than the critical value 
6.5, which means the core is static but unstable,
thus the gravitational collapse could occur with small perturbation.
Therefore, it is possible that B68 is slightly supercritical and 
could be undergoing slow contraction.
The most well-known collapsing core L1544 has similar size and mass to
B68 and its collapse velocity is $\gtrsim$ 0.1 \kms\ (Lai \& Crutcher 2000).
If B68 collapses with a velocity of the same order of magnitude, 
the collapse motion should have been detected with our
spectral resolution ($\sim$ 0.06\kms).
However, if B68 has just initiated collapse, which in the standard Shu 
inside-out scenario originates in the center, we may not be able to
spatially resolve the collapse with the beam size in our maps. 
With the spatial resolution of the IRAM 30m observations Lada et al.\ (2003) 
have suggested that the outer layers of the B68 core have small amplitude 
non-radial oscillations or pulsations about an equilibrium configuration. 
Higher spatial and spectral  resolutions   line  observations will be needed to
determine if B68 has initiated collapse deep in the center.

\subsection{Depletion and Chemical Differentiation in B68}

The observed depletion of molecular species in dense cold cloud cores is 
broadly consistent with time dependent chemical modeling (c.f. Bergin and 
Langer 1997). The well defined density structure (Bonnor Ebert sphere) of B68 
makes it a unique  core to study the chemical structure and evolution of dense 
starless cores. Bergin et al (2002) have found two orders of magnitude decrease 
in  the C$^{18}$O abundance  across the core, from low (A$_v$ $<$ 2mag)  
to high (A$_v$ $>$ 20mag) density regions. 
Even \n2h+, which is believed to be a 
non-depleting  molecular probe shows a decrease in abundance by factor of 2 
between the low density (A$_v$ $<$ 3mag) and high density (A$_v$ $>$ 20mag) 
regions. 
Di Francesco et al. (2002) found that  all the molecules (CO, CS, HCO$^+$,   
H$_2$CO, C$_3$H$_2$, \n2h+, \nh3)  have lower abundances in B68 than in other 
cloud cores.  However, only a few molecules have been mapped in 
detail, CO, CS, and \n2h+ (Bergin et al. 2002; Hotzel et al. 2002; 
Lada et al.\ 2003).  Our data
include fully sampled spectral line intensity maps of three additional 
species  \nh3, C$_3$H$_2$ and 
CCS. While NH$_3$ emission shows an enhancement near the extinction peak, 
CCS and C$_3$H$_2$ emissions show emission peaks away from the center with 
distinctly low emission at the extinction peak.
As seen in Figure 1, the CCS and C$_3$H$_2$ emission peaks appear well outside  
the brightest \n2h+ and \nh3 emission. The CCS and \c3h2 emissions seem to 
trace only the outer low density regions of the core.  Our 
results suggest that  CCS and C$_3$H$_2$ are severely depleted at the core 
center. Both species show a strong asymmetry  about the center roughly 
along the northeast to southwest direction. The CS emission (Lada et al.\ 2003) 
also shows somewhat similar asymmetry, 
and the intensity map of CCS emission is broadly 
consistent with that of CS. The strong CCS emission to the southeast of the 
core is coincident with the strongest CS emission.  
The relative intensities of their emissions at the extinction peak suggest 
that CCS is more severely depleted than CS.
Such difference between CCS and CS is consistent with the 
predictions of the  depletion models of 
Bergin and Langer (1997) that CCS depletion occurs before CS. We conclude that 
CCS, CS and \c3h2 do not trace the core center. 
But they trace  the  chemistry in the outer layers of B68. 
The chemical gradients   traced by these molecules may be  
the result of the local conditions such as the UV penetration. Indeed the 
extinction map (ALL2001) shows  a sharper boundary to the the east and 
northeast, 
coincident with the strongest CCS emission.  Lada et al.\ (2003) have suggested 
that B68 might have undergone interaction with the Loop I supernova bubble. 
Interestingly the center of this bubble is located to the  southwest of B68.  
It is not unlikely that some of the asymmetries in the chemical structure as 
traced by CCS and \c3h2  resulted from such an interaction. 

The NH$_3$ emission is resolved and not strongly centrally peaked as would 
be expected for the density structure of B68 core. 
This suggests that NH$_3$ too is somewhat depleted in the 
center. In Table 3 we compare the column densities for NH$_3$, CCS and 
C$_3$H$_2$ at two positions: at the NH$_3$
peak and the CCS peak.  The column densities for CCS and 
C$_3$H$_2$ have uncertainties of order 2--3
due to assumed T$_{ex}$. Nevertheless the 
column densities in Table 3 show, roughly,  that NH$_3$ has a shallower 
depletion (by about factor 2) than CCS and C$_3$H$_2$   
($>$ factor 3).  The large beam size of the maps does not resolve fully 
the emissions at the two positions (CCS and \nh3 peaks) and therefore, 
the CCS and \c3h2 column densities  at the \nh3 peak are rough upper limits. 
Furthermore, the excitation analysis for NH$_3$ is more reliable than 
for the others. (Using the hyperfine lines of NH$_3$ provides a good estimate 
of the optical depth). The 
\nh3 results are consistent with  depletion on an angular    scale and 
magnitude similar 
to that of \n2h+, as suggested by Bergin et al. (2002).

Since CCS is an early time molecule and \nh3 is abundant at a later stage,
the column density ratio of CCS and \nh3 has been suggested as a possible
indicator for cloud evolutionary time 
(Suzuki et al.\ 1992; Bergin \& Langer 1997).
As the evolutionary status of B68 has been established to be a likely  
pre-protostellar 
core on the verge of collapse, in Table 4 we  compare its N(CCS)/N(\nh3)
to two other  well-studied pre-protostellar cores, L1498 and L1544.
A comparison of the column densities and core sizes indicates that B68 has the 
lowest  CCS and \nh3   abundances, in agreement with the general conclusion of 
Di Franceso et al. (2002) that in B68 the molecular abundances are lower than 
those in dense clouds.
Among these three cores, L1544 is the most evolved core, because
the collapse motion has been directly observed (Tafalla et al.\ 1998;
Lai \& Crutcher 2000).  L1498 could be younger than B68, 
because CCS is more widely distributed in L1498 than in B68.
From the values of N(CCS)/N(NH$_3$) listed in Table 3,
B68 appears to be in an intermediate stage between L1498 and 
L1544 in core evolution.  

\section{Conclusion} 

B68 has been suggested to be a pre-protostellar core that can be described
as a  Bonnor-Ebert sphere at 16 K. 
Our DSN 70m maps show that B68 resembles the chemically differentiated
structure seen in other pre-protostellar cores;
the \nh3 peak coincides with the dust continuum peak,
whereas CCS and \c3h2 are offset from the \nh3 peak.
We derive a kinetic temperature $\sim$ 11 K in B68 from \nh3 (1,1) and (2,2) 
observations and turbulence of $\Delta v_{nth} = 0.3 \Delta v_{th}$.
Here we derive an empirical formulation for a modified 
Bonnor-Ebert sphere that includes turbulent, and magnetic pressures, 
and allows us to constrain the distance to 
B68 using the observed kinetic temperature and turbulence. We estimate the 
distance to B68 to be $\sim$ 95pc, on the near side of $\rho$ Oph,
with a corresponding core mass of $\sim$ 1.0 $M_{\odot}$. 
The sulfur species CCS and carbon chain molecule \c3h2 are heavily depleted 
in B68 and do not trace the dense interior region. We show some  
evidence for depletion of \nh3 at the core center roughly on 
a scale similar to that of \n2h+. 

\begin{acknowledgements}

We would like to thank the referee for valuable comments and suggestions.
This research was conducted at JPL, Caltech with support from NASA,
while SPL held a National Research Council Resident Research Associateship
at JPL.

\end{acknowledgements}

\clearpage

\begin{figure}
\plotone{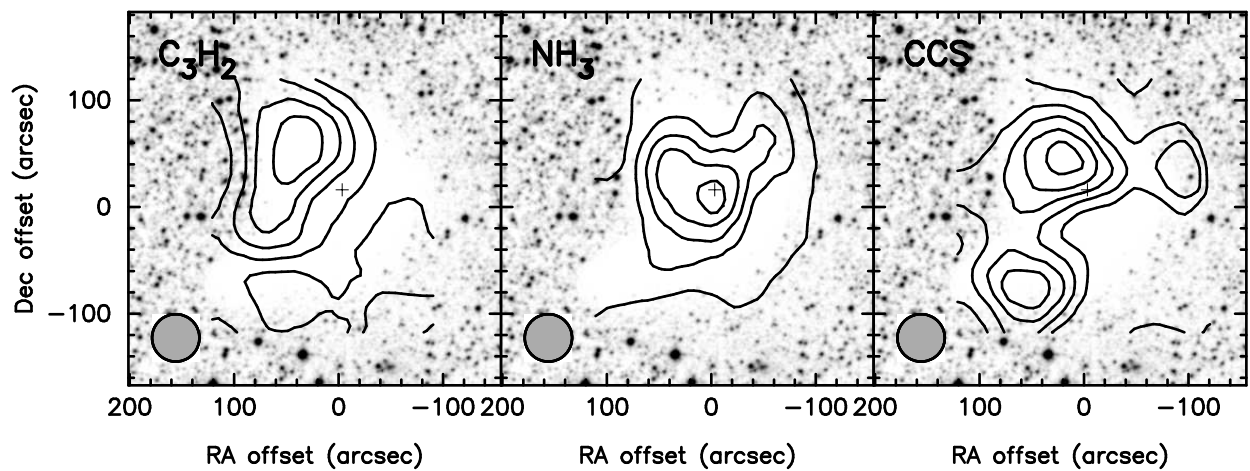}
\figcaption[Lai.fig1.ps]{DSN 70~m maps of \c3h2, \nh3, and CCS in B68 overlaid 
on the Palomar Optical Sky Survey image. The contours are 10, 30, 50,
70, and 90\% of the peak integrated antenna temperature, 
which are 0.23, 0.25, and 0.084 K$\cdot$\kms\
for \c3h2, \nh3, and CCS, respectively.
The (0,0) position corresponds to R.A.(1950)=17$^h$19$^m$36$^s$, 
Dec.(1950)=$-$23\deg47\arcmin13\arcsec.
The plus symbol indicates the extinction peak. }
\end{figure}

\begin{figure}
\plotone{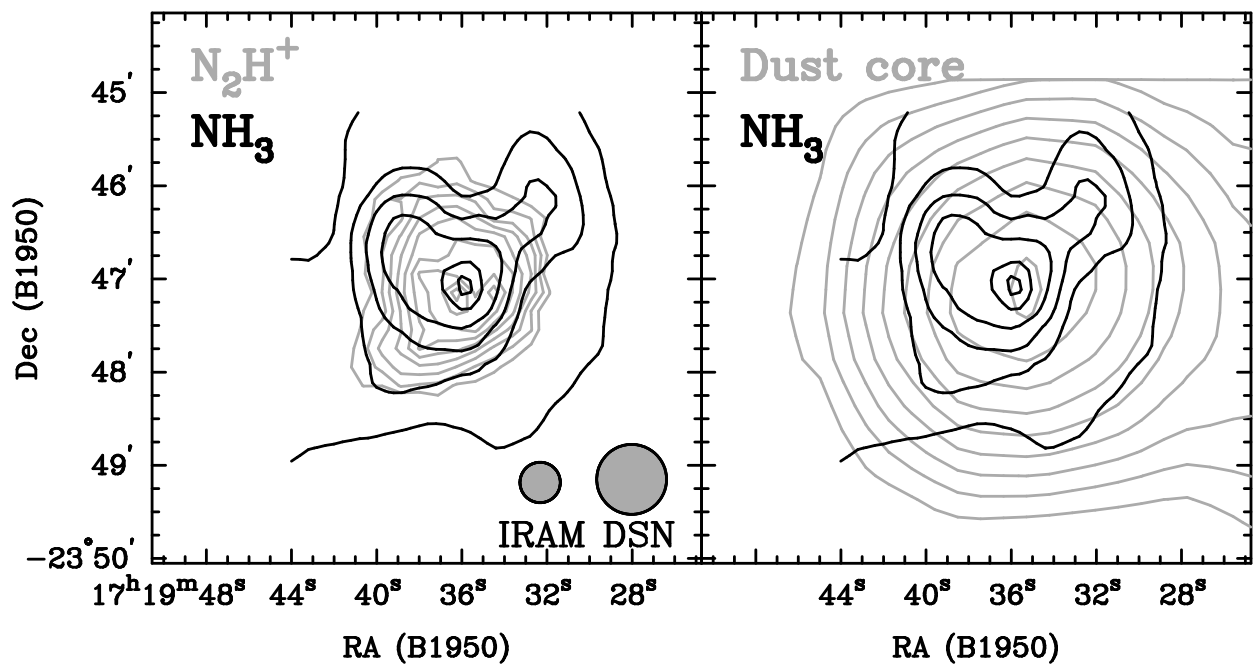}
\figcaption[Lai.fig2.ps]{Comparison of DSN 70m
\nh3 map with IRAM 30m N$_2$H$^+$ map (Left; Bergin et al. 2002)
and 200$\mu$m dust emission (Right; Langer \& Willacy 2001).}
\end{figure}

\begin{figure}
\plotone{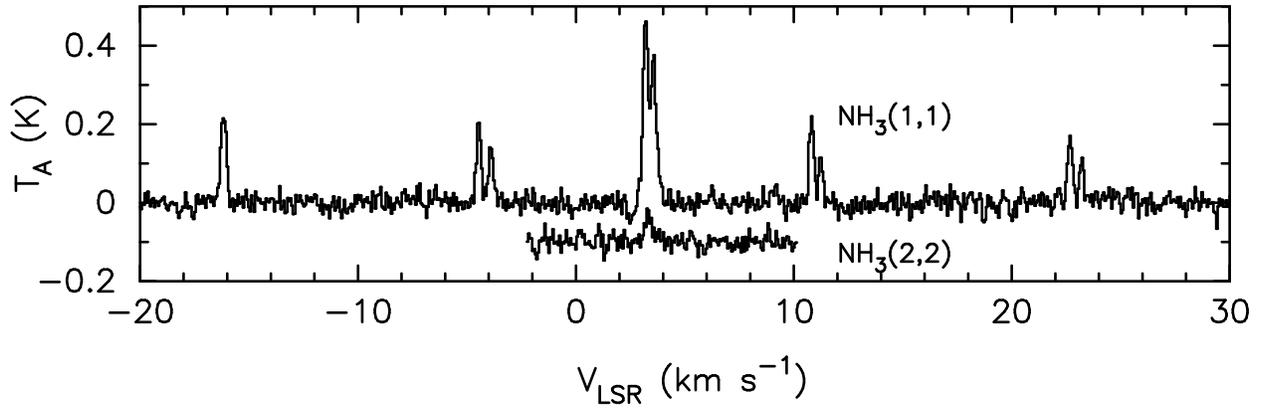}
\figcaption[Lai.fig3.ps]{DSN 34m \nh3 (1,1) and (2,2) spectra of B68.}
\end{figure}

\begin{figure}
\plotone{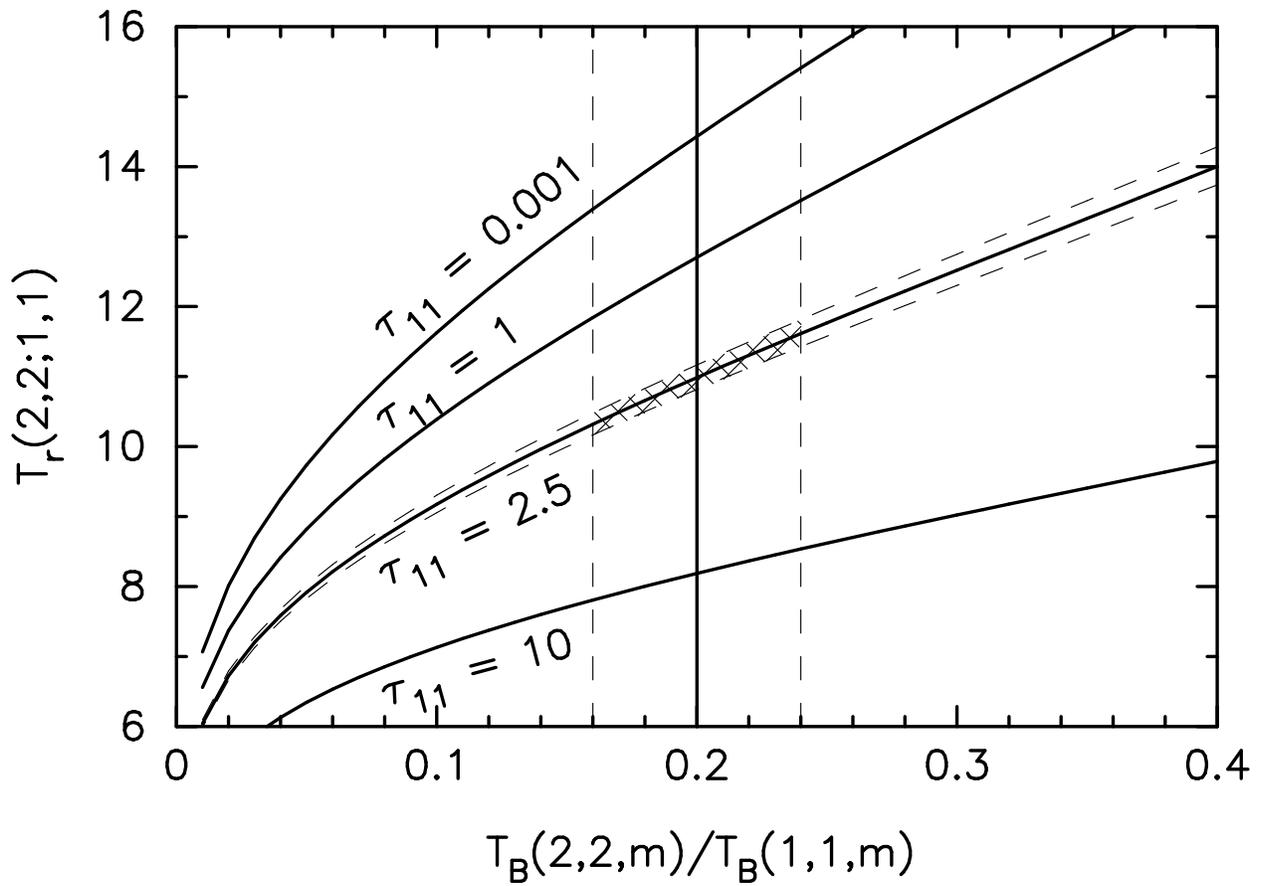}
\figcaption[Lai.fig4.ps]{Determination of the rotational Temperature, T$_r$.}
\end{figure}

\begin{figure}
\plotone{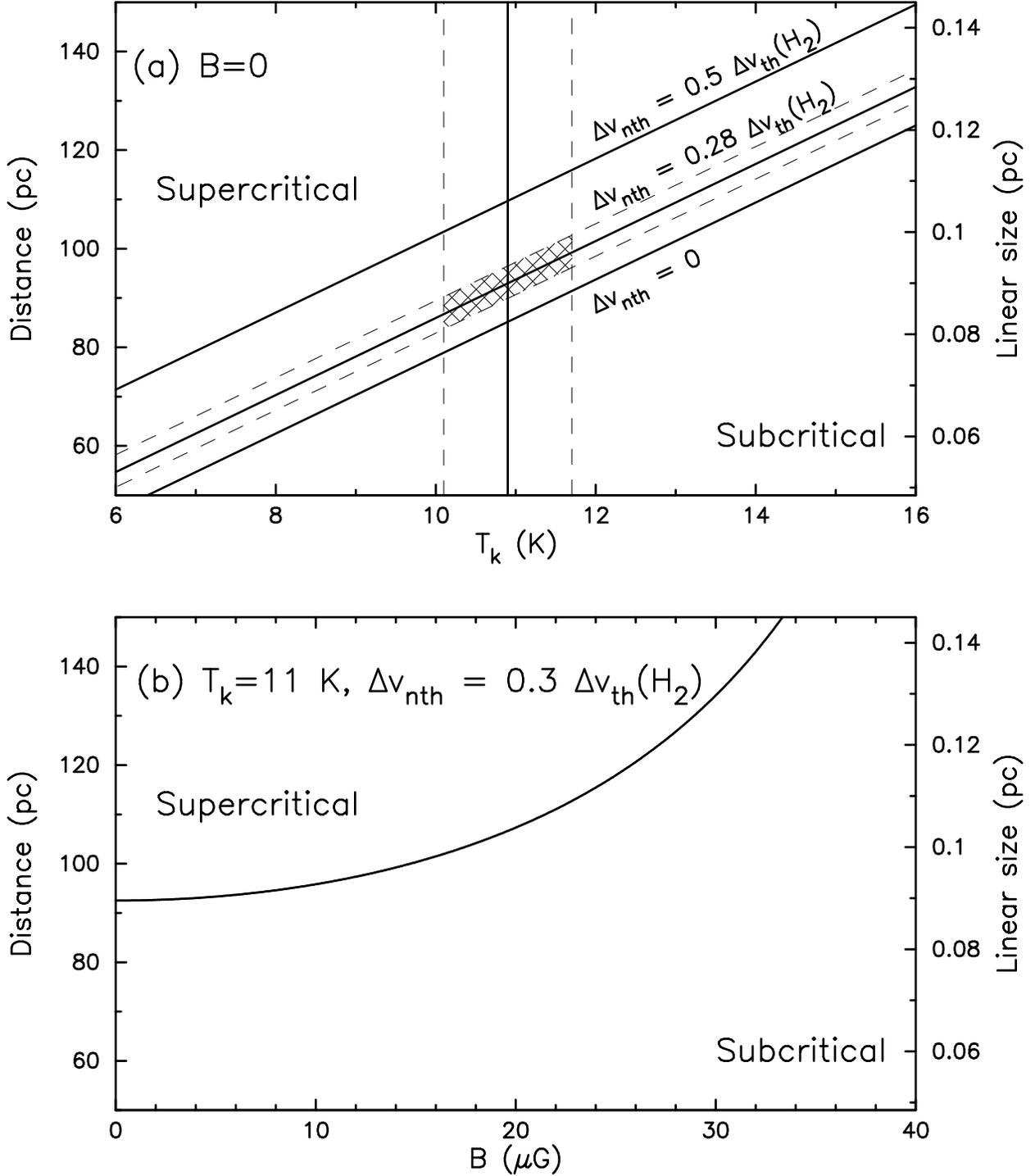}
\figcaption[Lai.fig5.ps]{(a) Distance (Left axis) constrained by 
temperature for magnetic field B=0 (see Eqs. 8 \& 10 in the text).
The turbulence levels are indicated.
The broken lines correspond to the uncertainty (1$\sigma$)
in the observed temperature and turbulence. The shaded area
represents the best estimates for the distance.
The corresponding linear size is given on the right axis.
(b) Distance constrained by magnetic fields.
Here we use the observed core temperature (11K) and
turbulence $\Delta v_{nth} = 0.3 \Delta v_{th}$.}
\end{figure}

\clearpage

\begin{table}
\begin{center}
\caption{Observations and Resolutions}
\vspace*{0.3cm}
\begin{tabular}{cccccc}
\hline
\hline
Molecule & Transition & Frequency & Antenna & HPBW & Velocity Resolution \\
	 &	      &	(MHz)	  & 	& (arcsec)  & (\kms) \\
\hline
\c3h2	& $J_{K-,K+} = 1_{1,0}\rightarrow1_{0,1} $ & 18343.145 & DSN 70 m & 45 
& 
0.080 \\
CCS	& $J_N=2_1\rightarrow1_0$ & 22344.033	& DSN 70 m & 45  & 0.066 \\
CCS	& $J_N=2_1\rightarrow1_0$ & 22344.033   & DSN 34 m & 90  & 0.066 \\
\nh3	& $J, K = 1,1 $ & 23694.495 & DSN 70 m & 45  & 0.062 \\
\nh3	& $J, K = 1,1 $ & 23694.495 & DSN 34 m & 90  & 0.062 \\
\nh3	& $J, K = 2,2 $ & 23722.633 & DSN 34 m & 90  & 0.062 \\
\hline
\hline
\end{tabular}
\end{center}
\end{table}

\begin{table}
\begin{center}
\caption{Linewidths of \nh3 and CCS}
\begin{tabular}{lcc}
\hline
\hline
& \nh3(1,1) & CCS\\
\hline
V$_{LSR}$ (\kms) & 3.35\e0.01 & 3.36\e0.02 \\
$\Delta v_{obs}$ (\kms) & 0.22\e0.01 & 0.17\e0.03 \\
$\Delta v_{th}$ (11K) & 0.17 & 0.094 \\
$\Delta v_{nth}$ & 0.14 \e0.01 & 0.14\e0.03 \\
\hline
\hline
\end{tabular}
\end{center}
\end{table}
\begin{table}
\begin{center}
\caption{Column densities at two positions}
\begin{tabular}{lccc}
\hline
\hline
Position & $N(NH_3)$ & $N(CCS)^a$ & $N(C_3H_2)^a$ \\
& ($10^{14}$ cm$^{-2}$) & ($10^{12}$ cm$^{-2}$) & ($10^{12}$ cm$^{-2}$) \\
\hline
NH$_3$ peak (0\arcmin,0\arcmin)& 1.9   & 1--1.5   & 0.8--3.3\\
CCS peak (0\farcm28,0\farcm7)  & 3.1   & 2.1--5.5 & 2.4--10 \\
\hline
\hline
\end{tabular}
\end{center}
$^a$ - the lower and upper limits of the column density for CCS and 
\c3h2  are correspond to the 
assumed $T_{ex}=5$ and 10 K, respectively. For \nh3, $T_{ex}$ derived 
from the spectral line data were used.
\end{table}

\begin{table}
\begin{center}
\caption{Comparison of L1498, B68, and L154}
\begin{tabular}{lcccc}
\hline
\hline
				 & L1498            & B68   & L1544  \\
\hline
$N(CCS)~(10^{12}$ cm$^{-2}$)     & 6.5$^{(1)}$      & 1--1.5     & 1--4$^{(3)}$ 
\\
$N(NH_3)~(10^{14}$ cm$^{-2}$)    & 8$^{(2)}$        & 2     & 10$^{(2)}$ \\
$\frac{N(CCS)}{N(NH_3)}$   	 & 0.008            & 0.005--0.0075 & 
0.001--0.004 \\
Size~(pc)			 & 0.07$\times$0.16 & 0.12  & 0.06$\times$0.12  
\\
$T_k$~(K)			 & 7--10$^{(1)}$    & 11    & 10$^{(2)}$ \\
$\Delta v_{nth}$~(km~s$^{-1}$)   &0.18--0.21$^{(4)}$& 0.14  & 0.25$^{(2)}$ \\
$\frac{E_{nth}}{E_{th}}$	 &0.12--0.13        & 0.09  & 0.32 \\
\hline
\hline
\end{tabular}
\end{center}
References -- (1) Wolkovitch et al.\ 1997. (2) Jijina, Myers, \& Adams 1999.
(3) Lai \& Crutcher 2000.  (4) Levin et al.\ 2001.
\end{table}

\end{document}